# RESPONSE TO "Time Modulations and Amplifications in the Axion Search Experiments"

S. Bertolucci [1], H. Fischer [2], S. Hofmann [3], M. Maroudas [4], Y.K. Semertzidis [5], K. Zioutas [4]

1) INFN, LNF, Italy

2) University of Freiburg, Freiburg 79104, Germany

3) Munich, Germany

4) University of Patras, Patras, Greece

5) CAPP / IBS and Department of Physics, KAIST, Daejeon 34141, Republic of Korea

Contact: zioutas@cern.ch

**ABSTRACT:**

In this short note we argue why the conclusions made by the authors of ref.[a] related to ref.[1] and in particular ref.[64] given therein, are incorrect. Nevertheless, we think that the advocated Anti-Quark Nuggets (AQNs) are good candidates for the dark Universe, and therefore they deserve further attention.

In arXiv:1908.04675 [astro-ph.CO] [a], X. Liang, A. Mead, Md Shahriar Rahim Siddiqui, L. Van Waerbeke and A. Zhitnitsky (LMSWZ), suggested a time dependent behaviour of axions originating from Anti-Quark Nuggets (AQNs) interaction with earth crust. LMSWZ arrived to conclusions contradicting ref.[1] and mainly ref.[64] cited in [a]. Here, we comment on the work by LMSWZ, because it is based on key assumptions incorrectly ascribed to [1] and in particular in [64] even explicitly excluded. While LMSWZ refer to axions, the authors of ref. [64] introduced generically a yet unidentified slow streaming invisible form of massive matter, which must interact with normal matter much stronger than the one expected for WIMPs or axions. In addition, the authors in [a] refer to the dark matter wind being obliged by ~60º to the ecliptic: ref.[64] concluded, observationally driven, on stream(s*)* in the ecliptic plane being temporally collinear with the line connecting the Sun and one planet (e.g. the Earth), for gravitational lensing to be possible. In ref.[1] it is explained how such a streaming configuration can result to the flux amplification downstream at 1 A.U. by factor up to ~$10^{11}$, which is to be compared with the derived amplification by LMSWZ of about $10^2$-$10^3$. Though, the authors consider a) ordinary dark matter wind, and b) a propagation being quasi orthogonal (~60º) to the ecliptic plane. We wish to stress that both assumptions are explicitly excluded in ref. [64].

In the following we reject the 3 key assumptions derived by LMSWZ in [a] (section "E. Gravitational lensing") for ref. [64], which are obvious misinterpretations to statements made in [64]:

1. *The deflection angle γ due to gravitational focusing is small, namely γ <<1*;
   This statement is wrong for large impact parameter b. Gravitational lensing by the Sun at 1 A.U. and small incident particle velocities requires an accordingly large impact parameter b ($\gamma \sim M(b)/v^2 b$). This global constraint on γ is apparently inapplicable for the only relevant case in ref. [64], namely that of the dynamic ionosphere. The same reasoning answers also our next objection.

2. *The DM flux has no dispersion in velocity*;
   Citing ref. [64] (section 2): "… it is reasonable to assume that constituents of any kind of invisible massive matter having a wide velocity spectrum around 300 km/s can undergo gravitational focusing towards the Sun, if one of the planets considered in this work, i.e., Mercury, Venus and/or Earth, enters inside such a stream towards the Sun." See also the reasoning given before for **1**.

3. *The DM particles are non-interacting and can pass through opaque objects such as the Sun and Planets;*
   Citing ref. [64], section 3: "a. Slow moving invisible (streaming) matter of galactic/cosmic origin, whatever its eventual properties, interacts somehow with the Sun…."

Furthermore, based on the before mentioned, incorrectly reproduced key assumptions LMSWZ draw the conclusion: **"... that the ideas advocated in Refs. [63, 64] generally do not apply to DM particles."**
We comment on this conclusion with the very first working hypothesis given in ref.[64], which apparently has not been noticed by the authors of [a]: "…we refer to generic dark candidate constituents as invisible massive matter, in order to distinguish them from ordinary dark matter".

We think, after all this, it is redundant to elaborate on more points raised by LMSWZ.

**Following the new version of [a]:**

The authors LMSWZ uploaded recently a 2nd version of [a], which, however, does not address satisfactorily our main points given above. Therefore, we add below some more comments (in italic), referring in the following to version-2 of [a]. Thus:

**A.** From [a], page 9, left:

"… on Fig. 4. Such large deflection angle are impossible to obtain …, consequently, calculation in Refs. [63, 64] is no longer applied..."
*The configuration of Fig. 4 was never considered as an option in [64], i.e., to have simultaneously gravitational lensing and deflection by 62º.*
"…Next, assumption 2 (=The DM flux is collinear) is violated because AQNs (and in fact most DM candidates) have a large velocity dispersion ∼ 110 km s$^{-1}$..."
*A large velocity dispersion is not a measure of being collinear.*
"…Thus, the enhancement of DM flux by gravitational lensing as advocated in Refs. [63, 64] has a very narrow window for applicability..."
*Yes, but for the incorrect assumption, i.e. of not being collinear.*
"… even for conventional DM candidates like WIMPs, for which both assumptions becomes invalid in the present framework of SHM."
*We repeat that conventional DM candidates are explicitly excluded in [64], and, the observationally derived conclusions there cannot be within the SHM.*

**B.** From [a], page 9, right, about the enhancement up to $10^{11}$ given in [1].

*In ref. [63] of [a] it was calculated that even Jupiter can give a flux enhancement by factor up to $10^6$. Then it is reasonable to expect that the Sun should be even a better performing gravitational lens. In addition, citing https://arxiv.org/abs/1703.01436: "In the ideal case of perfect alignment Stream-Sun-Earth, the axion flux enhancement of the stream can be very large (~$10^6$) or even much more …". Notably, such enhancements are missing in both versions of [a], most probably because of unrelated assumptions (see next).*

**C.** From [a], page 9, right:

"…if assumption 2 (=The DM flux is collinear) is also enforced (θ = 0º, σ = 1 km s$^{-1}$)…"
*This is wrong and it explains eventually all objections raised by the authors LMSWZ. Namely, the DM wind is not a usual stream as those assumed in [1] and [63,64]. Because, DM particles are moving randomly in our Galaxy with a broad Maxwell distribution at ~270 km/s (see, for example, the plot #8 in https://slideplayer.com/slide/7719764/ ).*